\documentclass[12pt]{article}

\usepackage[T1]{fontenc} \usepackage[dvips]{graphics}

\begin{document}

\title{Null Cones in Schwarzschild Geometry}

\author{Thomas P.  Kling, Ezra T.  Newman,\\Department of Physics and Astronomy\\University of
Pittsburgh\\ Pittsburgh, Pa, 15260}

\date{\today}

\maketitle

\begin{abstract} In this work we investigate aspects of light cones in a Schwarzschild geometry, making
connections to gravitational lensing theory and to a new approach to General Relativity, the Null
Surface Formulation.  By integrating the null geodesics of our model, we obtain the light cone from
every space-time point.  We study three applications of the light cones.  First, by taking the
intersection of the light cone from each point in the space-time with null infinity, we obtain the
light cone cut function, a four parameter family of cuts of null infinity, which is central to the Null
Surface Formulation.  We examine the singularity structure of the cut function.  Second, we give the
exact gravitational lens equations, and their specialization to the Einstein Ring.  Third, as an
application of the cut function, we show that the recently introduced coordinate system, the
``pseudo-Minkowski'' coordinates, are a valid coordinate system for the space-time.  \end{abstract}

\section{Introduction}

The purpose of this work is to develop a simple model space-time in which we study gravitational
lensing and the light cone cuts of null infinity.  These cuts are central to a recent reformulation of
General Relativity known as the Null Surface Formulation~\cite{simo1,simo2}.  The model we consider
consists of a Schwarzschild exterior region surrounding a spherically symmetric, static, constant
density dust star.

The Null Surface Formulation makes explicit use of future null infinity, denoted by $\cal{I}^+$, which
has the topology of $\mbox{\rm\bf R} \times S^2$.  In an asymptotically simple space-time, $\cal{I}^+$
represents the future endpoints of all null geodesics and can be added as a boundary to the physical
space-time through a process of conformal compactification~\cite{Penrose}.  The standard coordinates on
$\cal{I}^+$ are the Bondi coordinates $(u, \theta, \phi)$, where $u$ labels the $\mbox{\rm\bf R}$ part,
and $\theta$ and $\phi$ label the sphere.

On $\cal{I}^+$, a light cone cut is the intersection of the light cone from a particular point in the
interior with future null infinity.  In Bondi coordinates on $\cal{I}^+$, the light cone cuts are given
by a cut function,

\begin{equation} u = Z(x^a_\circ, \theta, \phi), \label{Z}\end{equation}

\noindent where for a fixed initial point, $x^a_\circ$, the cut is a deformed sphere, possibly with
self-intersections and singularities.  The crux of the Null Surface Formulation is that from the
knowledge of the light cone cuts, Eq.~(\ref{Z}), one can reconstruct all of the conformal information
of the space-time.

The current paper explores the light cones from an arbitrary point in a Schwarzschild space-time
surrounding a constant density, dust star.  By integrating the null geodesics in an inverse radial
coordinate $l~=~1/\sqrt{2}r$, we obtain parametric expressions for the future and past light cones of
an initial point, $x^a_\circ$, in terms of a ``distance'' $l$, and two ``directional'' parameters which
span the sphere of null directions at $x^a_\circ$.  The intersection of the future light cone of an
arbitrary initial point with null infinity gives the light cone cuts.  In Section 3, we study the
singularity structure of the cuts, explaining how the singularities are related to the formation of
conjugate points on the light cone.  We show in Section 4 that the equations for the past light cone
obtained in Section 2 are, in fact, examples of exact lens equations.  As a special case, we give the
exact formula for the observation angle for an Einstein Ring.  In the final section, we show that the
``pseudo-Minkowski'' coordinates, defined in \cite{simo3}, form a valid coordinate system for the
entire space-time.

\section{Integrating the Null Geodesics}

To begin, we integrate the null geodesics of a Schwarzschild space-time with an interior constant
density matter region.  Since we eventually consider each geodesic's limiting endpoint at $\cal{I}^+$
in order to obtain a cut function, we integrate the null geodesics using a conformal Schwarzschild
metric which is regular at null infinity.  The integration is performed using a ``radial'' parameter
$l~=~1/(\sqrt{2}r)$, so that $l = 0$ corresponds to the point at null infinity, while a finite $l >0$
will be a point in the interior.

The general form for a static, spherically symmetric metric with signature $(+, -, -, -)$ is given by

\begin{equation} ds^2 = f(r) dt^2 - h(r) dr^2 - r^2d\Omega^2, \label{realg} \end{equation}

\noindent where $r^2d\Omega^2 = r^2 d\theta^2 + r^2 \sin^2\theta d\phi^2$ is the line element on the
sphere.  In our model, the functions $f(r)$ and $h(r)$ will be continuous, piecewise smooth functions
for an exterior Schwarzschild region and an interior constant density dust solution with a radius $R$:

\[ f(r) = \left(\frac{3}{2}\left(1 - \frac{2M}{R}\right)^{\frac{1}{2}} - \frac{1}{2}\left(1-\frac{2 M
r^2}{R^3}\right)^{\frac{1}{2}}\right)^2 \equiv f_{int}, \quad r <R, \]

\[ f(r) = 1 - \frac{2M}{r}\equiv f_{ext}, \quad\quad r > R, \]

\[ h(r) = \frac{1}{1 - \frac{2M r^2}{R^3}} \equiv h_{int}, \quad\quad r<R,\]

\begin{equation} h(r) = \frac{1}{1 - \frac{2M}{r}} \equiv h_{ext}, \quad\quad r>R.  \label{fandh}
\end{equation}

\noindent We assume that the radius of the interior region extends beyond the $r = 3M$ unstable
circular orbit for null geodesics, ensuring that the space-time is asymptotically simple.  Working in a
retarded null coordinate, $u = t - \int dr \sqrt{h(r)/f(r)}$, and the inverted radial coordinate,
$l~=~1/(\sqrt{2}r)$, a conformally rescaled version of the metric, Eq.~(\ref{realg}), which is regular
at null infinity, is

\begin{equation} d\hat{s}^2 = 4l^2~f(l)~du^2 - 4\sqrt{k(l)}du~dl - d\Omega^2, \label{confrmg}
\end{equation}

\noindent where $r$ is replaced in terms of the variable $l$ and $k(l)$ is given by

\[ k(l) = f(l)~h(l).  \]

It is convenient to integrate the null geodesics first in a plane, and then to use the spherical
symmetry to rotate the solution to an arbitrary orientation.  The restriction we employ is to take a
particular initial point, ${\tilde{x}}^a_\circ$, lying on the $-\hat z$ axis,

\[ {\tilde{x}}^a_\circ = (u_\circ, l_\circ, \tilde{\theta}_\circ = \pi, \tilde{\phi}_\circ = 0), \]

\noindent and the geodesics as lying in the $\hat x$-$\hat z$ plane.  To define the $\hat x$-$\hat z$
plane, one usually allows $\theta$ to range from $0$ to $\pi$ and has $\phi = 0$ or $\pi$.  In order to
facilitate our discussion, we will not use this convention.  Instead, we require that $\phi = 0$ and
allow a variable, $\Theta$, to range from $0$ to $2\pi$.  In terms of the variables $(u, l, \Theta)$, a
Lagrangian corresponding to the conformal metric is

\begin{equation} {\cal L} = 2l^2~f(l)~{\dot u}^2 -2\sqrt{k(l)}~{\dot u}~{\dot l} - \frac{1}{2}{\dot
\Theta}^2, \label{3Lag} \end{equation}

\noindent where dots indicate derivatives with respect to an affine parameter $\tau$.  The geodesic
equations are

\[ \frac{d}{d\tau}(2l^2~f(l)~\dot u - \sqrt{k(l)}~\dot l) = 0,\] \[ \frac{d}{d\tau}(-\sqrt{k(l)}~\dot
u) - (2l~f(l)~\dot u^2 + l^2~f'(l)~\dot u^2) + \frac{\dot u~\dot l}{2\sqrt{k(l)}}(k'(l)) = 0,\]
\begin{equation} \frac{d}{d\tau} \dot\Theta = 0 ,\label{e-l-plane} \end{equation}

\noindent with primes denoting derivatives with respect to $l$.  To solve for null geodesics, we impose
the null condition on the Lagrangian,

\begin{equation} {\cal L} = 2l^2~f(l)~{\dot u}^2 -2\sqrt{k(l)}~{\dot u}~{\dot l} - \frac{1}{2}{\dot
\Theta}^2 = 0, \end{equation}

\noindent and use this equation and the first and third geodesic equations as independent equations for
$(u, l, \Theta)$.  After finding some trivial first integrals and rearranging, independent equations
for null geodesics can be written as

\[ \dot u = {{1 + \sqrt{k(l)}~\dot l}\over{2 l^2~f(l)}},\] \[ \dot l = \pm\sqrt{{{1 -
b^2~l^2~f(l)}\over{k(l)}}} \equiv \pm \sqrt{A(l)},\] \begin{equation} \dot \Theta =
b.\label{eqnmotion-plane} \end{equation}

\noindent In these equations the sign of $\dot l$ indicates whether the geodesic is incoming (positive)
or outgoing (negative), and the parameter $b$ is a free integration parameter labeling the initial
direction of the geodesic.  In Appendix A, we show how the parameter $b$ is related to the initial
direction of the null geodesic.

In integrating the equations, geodesics which are initially outgoing will have $b$ values ranging from
zero, corresponding to a ray traveling radially outward, to some maximum value $b_m$, for which the
geodesic is initially tangent to a sphere of radius $r_\circ = 1/(\sqrt{2}l_\circ)$.  The value of
$b_m$ is the value of $b$ for which $\dot l =0$ at the point $l = l_\circ$:

\begin{equation} b_m = \frac{1}{l_\circ\sqrt{f(l_\circ)}}.  \label{bmeqn} \end{equation}

\noindent For geodesics which are initially incoming, the range in $b$ will be from $b_m$ back down to
zero.  The value of $l$ will increase until reaching some maximum value, $l = l_b$, which is a minimum
$r$ value.  The value of $l_b$ is the single (real) root of the equation, $\dot l = 0$, or the root of:

\begin{equation} 1 - b^2~l_b^2~f(l_b) = 0.  \label{lmax} \end{equation}

\noindent At $l_b$, $\dot l$ changes sign, and the geodesics will head out to infinity.

It is convenient to reparametrize the equations of motion, Eqs.~(\ref{eqnmotion-plane}), using $l$
instead of the affine parameter, and to express the solution to the null geodesic equations in terms of
integrals over $l$.  In terms of these integrals, geodesics on the light cone connecting the initial
point, ${\tilde {x}}_\circ^a=(u_\circ,l_\circ,\theta_\circ~=~\pi, \phi_\circ~=~0)$, with the final
point, $x^a=(u,~l,~\Theta,~\phi~=~0)$, are given by

\[ u = u(u_\circ, l_\circ, l, b) = u_\circ + (-1)^{\epsilon} \int_{l_\circ}^l~dl'~\left(\pm {{1 \pm
\sqrt{k(l')~A(l')}}\over{\sqrt{A(l')}(2l'^2~f(l'))}}\right), \]

\[\Theta = \Theta (l_\circ, l, b) =\pi - \int_{l_\circ}^l~dl'~\left(\pm \frac {b}{\sqrt{A(l')}}\right),
\] \begin{equation} \phi = 0, \label{lcplane} \end{equation}

\noindent where the integrals are defined piecewise over the various segments in $f$ and $k$, and the
appropriate signs are chosen if the geodesics are incoming ($+$) or outgoing ($-$).  The future light
cone is given by $\epsilon = 0$, and the past light cone by $\epsilon = 1$.

The full solution to the null geodesic equations is obtained by performing a rigid rotation of the
spatial plane defined by the path of this geodesic and the $\hat z$ axis, to an arbitrary orientation.
Due to spherical symmetry, the light cone for an arbitrary point will be axially symmetric about the
spatial line connecting that point and the spatial origin (defined by $r = 0$).  We will call the angle
of revolution about the axis of symmetry $\gamma$; the angle $\gamma$ essentially defines an
orientation of the spatial plane containing the geodesic and the axis of symmetry.  (In the case that
the initial point lies on the $-\hat z$ axis, $\gamma$ is simply the angle $\phi$.)  To obtain the full
light cone from an arbitrary initial point, we need to rotate the solution that is symmetric about the
$\hat z$ axis over to the axis of symmetry defined by the arbitrary point.  We can think of this as
rotating the particular initial point ${\tilde
x}_\circ^a~=~(u_\circ,~l_\circ,~\theta_\circ~=~\pi,~\phi_\circ~=~0)$ to the general initial point
$x^a_\circ~=~(u_\circ,~l_\circ,~\theta_\circ,~\phi_\circ)$ and allowing the orientation parameter
$\gamma$ to take any value between $0$ and $2\pi$.  This rotation is explicitly performed in Appendix
B.

It is often convenient to use complex stereographic coordinates, defined by

\begin{equation} \zeta = \cot{\frac{\theta}{2}}e^{i\phi}, \quad\quad \bar\zeta =
\cot{\frac{\theta}{2}}e^{-i\phi}, \label{defzeta} \end{equation}

\noindent instead of the angular coordinates $\theta$ and $\phi$.  Throughout this paper we freely
switch back and forth between the two coordinate systems.  An arbitrary initial point is given in
stereographic coordinates by $x^a_\circ~=~(u_\circ,~l_\circ,~\zeta_\circ,~\bar\zeta_\circ)$.

The full light cone, obtained by using the rotation of the solutions of the null geodesic equations in
the $\hat x$-$\hat z$ plane to an arbitrary initial point and orientation given in Appendix B, are
expressed parametrically in terms of the parameter $l$ as

\[ u = u(u_\circ, l_\circ, l, b) = u_\circ + (-1)^{\epsilon} \int_{l_\circ}^l~dl'~\left(\pm {{1 \pm
\sqrt{k(l')~A(l')}}\over{\sqrt{A(l')}(2l'^2~f(l'))}}\right), \]

\[ \zeta(l_\circ, \zeta_\circ, \bar\zeta_\circ, l, b, \gamma) = \left(\frac{\zeta_\circ}{\bar
\zeta_\circ}\right)^{\frac{1}{2}}\left(\frac{e^{\frac{i}{2}\gamma} \cot{\frac{\Theta(l, l_\circ,
b)}{2}} + \sqrt{\zeta_\circ\bar\zeta_\circ} e^{-\frac{i}{2}\gamma}}{-\sqrt{\zeta_\circ\bar\zeta_\circ}
e^{\frac{i}{2}\gamma} \cot{\frac{\Theta(l, l_\circ, b)}{2}} + e^{-\frac{i}{2}\gamma}}\right), \]

\begin{equation} \bar\zeta(l_\circ, \zeta_\circ, \bar\zeta_\circ, l, b, \gamma) =
\left(\frac{\bar\zeta_\circ}{\zeta_\circ}\right)^{\frac{1}{2}}\left(\frac{e^{-\frac{i}{2}\gamma }
\cot{\frac{\Theta(l, l_\circ, b)}{2}} + \sqrt{\zeta_\circ\bar\zeta_\circ}
e^{\frac{i}{2}\gamma}}{-\sqrt{\zeta_\circ\bar\zeta_\circ} e^{-\frac{i}{2}\gamma} \cot{\frac{\Theta(l,
l_\circ, b)}{2}} + e^{\frac{i}{2}\gamma}}\right), \label{lightcone}\end{equation}

\noindent where $\Theta(l, l_\circ, b)$ is given by the integral in Eq.~(\ref{lcplane}),

\begin{equation} \Theta = \Theta (l_\circ, l, b) =\pi - \int_{l_\circ}^l~dl'~\left(\pm \frac
{b}{\sqrt{A(l')}}\right), \label{Theta}\end{equation}

\noindent and the convention for $\epsilon$ and signs are taken as before.

Eqs.~(\ref{lightcone}) represent the entire future and past light cones for an arbitrary point in the
space-time.  The light cones are given in terms of two parameters, $(b,~\gamma)$, which span the sphere
of initial null directions at the initial point $x^a_\circ$.  We note that the $u$ coordinate does not
depend on $\gamma$ due to the axial symmetry of the light cone.

\section{Light Cone Cuts and their Singularities}

\begin{figure} \begin{center} \scalebox{0.85}{\includegraphics{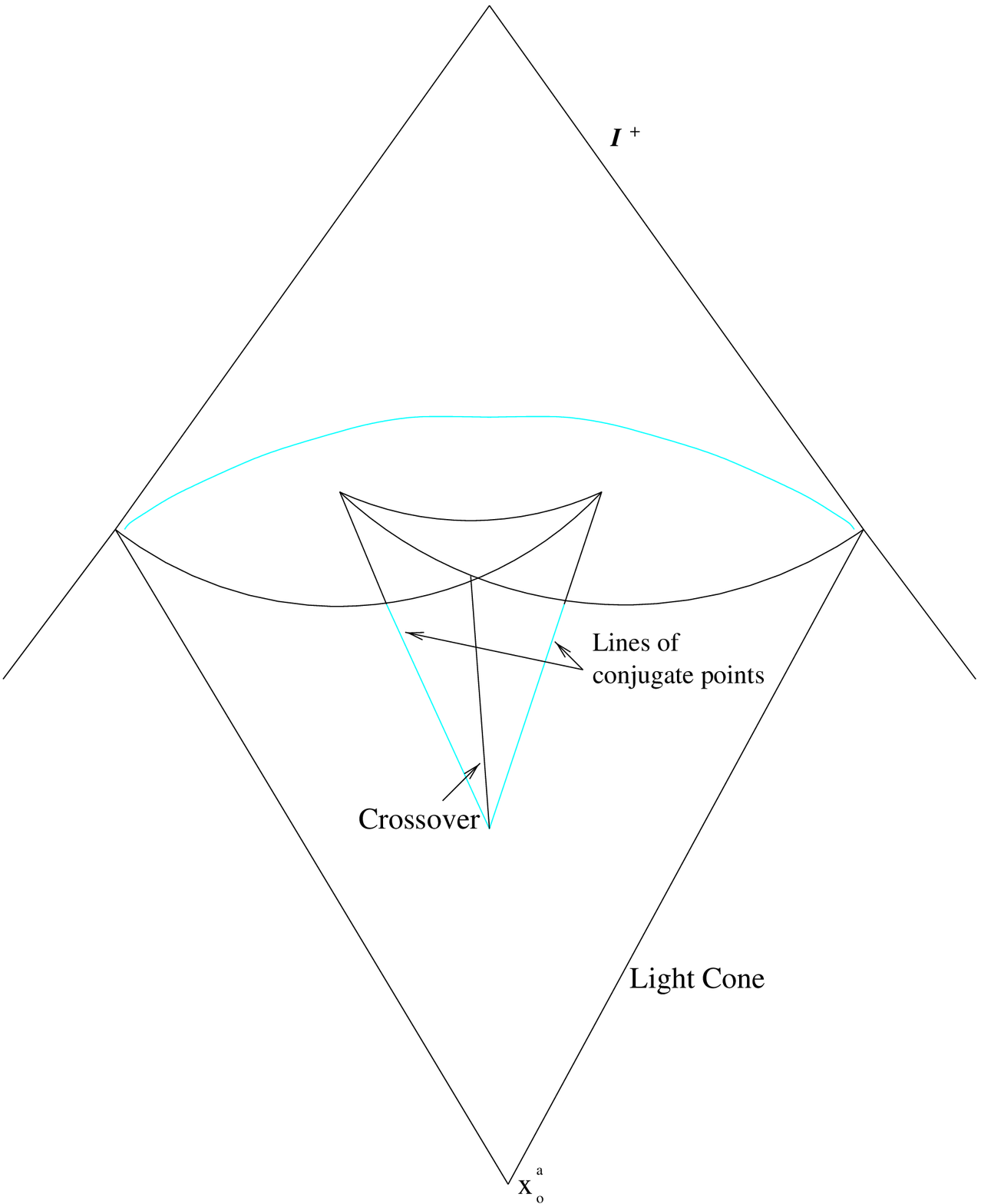}} \caption{The light cone
of a point which has formed conjugate points in its future.  The crossover line represents points in
the space-time where Einstein Rings are observed, while the line of caustics are stable singularities
conjugate to the initial point.}  \label{lightconepic} \end{center} \end{figure}

While the light cone from an arbitrary point in Minkowski space is always smooth, light cones in an
asymptotically simple space-time have, in general, self-intersections and several different kinds of
singularities.  These singularities are directly related to the formation of conjugate points along
null geodesics.  A pictorial representation of a light cone with singularities is given in
Figure~\ref{lightconepic}.  Since the light cone cut function is the intersection of the future light
cone with null infinity, it inherits the singularity structure of the light cone.  In this section, we
study the cut function in our model as a representative example of the Null Surface Formulation.

A parametric version of the cut function is obtained by setting the values of $l$ and $\epsilon$ to
zero in Eqs.~(\ref{lightcone}):

\begin{equation} u_\infty(u_\circ, l_\circ, b) = u_\circ + \int_{l_\circ}^0~dl'~\left(\pm {{1 \pm
\sqrt{k(l')~A(l')}}\over{\sqrt{A(l')}(2l'^2~f(l'))}}\right), \label{uatscri} \end{equation}

\begin{equation} \zeta_\infty(l_\circ, \zeta_\circ, \bar\zeta_\circ, b, \gamma) =
\left(\frac{\zeta_\circ}{\bar \zeta_\circ}\right)^{\frac{1}{2}}\left(\frac{e^{\frac{i}{2}\gamma}
\cot{\frac{\Theta(l_\circ, b)}{2}} + \sqrt{\zeta_\circ\bar\zeta_\circ}
e^{-\frac{i}{2}\gamma}}{-\sqrt{\zeta_\circ\bar\zeta_\circ} e^{\frac{i}{2}\gamma}
\cot{\frac{\Theta(l_\circ, b)}{2}} + e^{-\frac{i}{2}\gamma}}\right), \label{zetaatscri}\end{equation}

\begin{equation} \bar\zeta_\infty(l_\circ, \zeta_\circ, \bar\zeta_\circ, b, \gamma) =
\left(\frac{\bar\zeta_\circ}{\zeta_\circ}\right)^{\frac{1}{2}}\left(\frac{e^{-\frac{i}{2}\gamma }
\cot{\frac{\Theta(l_\circ, b)}{2}} + \sqrt{\zeta_\circ\bar\zeta_\circ}
e^{\frac{i}{2}\gamma}}{-\sqrt{\zeta_\circ\bar\zeta_\circ} e^{\frac{-i}{2}\gamma}
\cot{\frac{\Theta(l_\circ, b)}{2}} + e^{\frac{i}{2}\gamma}}\right), \label{barzetaatscri}\end{equation}

\noindent where the function $\Theta(l_\circ, b)$ is given by

\begin{equation} \Theta(l_\circ, b) = \pi - \int_{l_\circ}^0~dl'~\left(\pm \frac
{b}{\sqrt{A(l')}}\right).  \label{Theta1}\end{equation}

\noindent If it were possible to invert the pair of equations, Eqs.~(\ref{zetaatscri}) and
(\ref{barzetaatscri}), for $b$ and $\gamma$, obtaining the functions

\begin{equation} \gamma = G(\zeta_\infty, \bar\zeta_\infty, \zeta_\circ, \bar\zeta_\circ, l_\circ),
\end{equation}

\noindent and

\begin{equation} b = B(\zeta_\infty, \bar\zeta_\infty, \zeta_\circ, \bar\zeta_\circ, l_\circ),
\label{B} \end{equation}

\noindent one could produce the full cut function in the form of Eq.~(\ref{Z}),

\[ u = Z(x^a_\circ, \zeta, \bar\zeta), \]

\noindent by inserting the solution for $b$ in terms of $(\zeta, \bar\zeta)$ from Eq.~(\ref{B}) into
the $u_\infty$ solution, Eq.~(\ref{uatscri}).

While $\zeta_\infty (l_\circ,~\zeta_\circ,~\bar\zeta_\circ,~b,~\gamma)$, and $\bar\zeta_\infty
(l_\circ,~\zeta_\circ,~\bar\zeta_\circ,~b,~\gamma)$ are single valued in $b$, they will not, in
general, have unique inverses---more than one initial direction acquires the same value of $\zeta$ or
$\bar\zeta$ at $\cal{I}+$.  This implies that there are singularities in the cut function itself and
that the global inversion of Eqs.~(\ref{zetaatscri})~and~(\ref{barzetaatscri}) for $b$ and $\gamma$
will be impossible.  In such a case, we will not be able to find an explicit cut function in the form
of Eq.~(\ref{Z}), but will be forced to work with the cut function in a parametric form.

In a sense, singularities in the light cone cuts are places where the Null Surface Formulation
undergoes technical difficulties---a natural coordinate system used in the theory is not well defined
at these points.  We now believe that these difficulties can be overcome by using a particular
parametric representation the cut function.  A primary interest here is to study the singularities in
the cut function.

There is a complete classification of the stable singularities of the cut function which can be applied
to our model, due to Arnol'd and his collaborators~\cite{Arnold1,Arnold2}.  A stable singularity is one
which does not disappear under small perturbations.  For two dimensional surfaces, such as the light
cone cuts obtained by fixing the initial point in the cut function, there are only two types of stable
singularities.  These are the cusp ridge and the swallowtail.  Due to high level of symmetry in our
model, any cut function must be axially symmetric.  This means that, although the cut function is a two
dimensional surface, it can be represented by a one dimensional curve whose revolution about some axis
gives the cut function.  For a one dimensional curve, the only stable singularity is a cusp, which
implies that we will not see swallowtail singularities in our model.

\begin{figure}[!hp] \begin{center} \scalebox{1.0}{\includegraphics{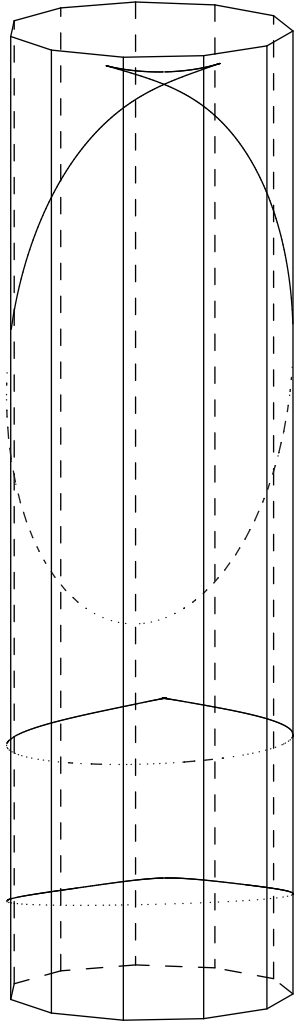}} \caption{Three cut
functions with one dimension suppressed ($S \times \mbox{\rm\bf R}$).  As the initial radial position
moves away from the spatial center smooth cuts give way to cuts with singularities.}  \label{cut}
\end{center} \end{figure}

\begin{figure} \begin{center} \scalebox{1.0}{\includegraphics{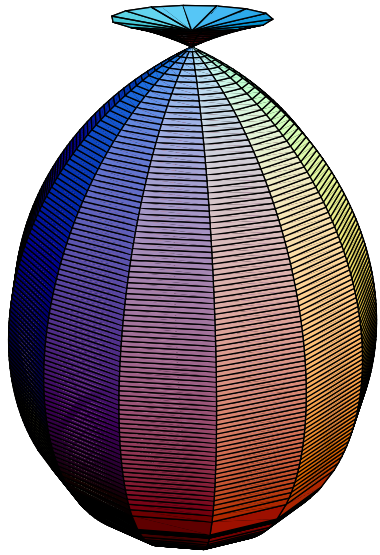}} \caption{A cusp ridge
appears on the light cone cut from an initial point sufficiently far away from the center of the
space-time.  Near $\cal{I}^+$, the intersection of the light cone of this initial point with a
time-like surface would look similar to this cut, except that the ``umbrella'' at the top of the cut
would be inside the main body of the wavefront.}  \label{3dcusp} \end{center} \end{figure}

In Figure \ref{cut}, we give plots of three cut functions, suppressing the axially symmetric dimension,
for three different values of the initial radial position,~$l_\circ$.  These figures show that
singularities appear as the initial point moves away from the spatial center of the space-time.  A
smooth cut of null infinity, corresponding to a cut from the light cone of a point close to the center
of the space-time, will be a smooth sphere-like surface.  Due to the axial symmetry, a cut with
singularities will have a circular cusp ridge and a single crossover point.  Figure \ref{3dcusp} gives
a pictorial representations of a singular cut.

Because the cusp ridge singularity in our cut function is stable, it represents a generic possibility
for a cut function in a general space-time.  Specifically, this singularity will remain if one makes a
small perturbation of the metric away from a Schwarzschild metric.  The crossover point along the cut
function is an unstable singularity, arising from the high degree of symmetry in the Schwarzschild
case, and is directly related to Einstein's Rings, the astronomical phenomenon where a spherical lens
causes lensing in a uniform circular shape~\cite{Ehlers}.  We discuss the rings in the next section on
gravitational lensing.

The singularities in the cut function represent points which are conjugate to the initial point
$x^a_\circ$.  For a fixed $x^a_\circ$, we can think of the parametric equations for the cut function,
Eqs.~(\ref{uatscri}),~(\ref{zetaatscri}),~and~(\ref{barzetaatscri}), as a map between the initial
directions of the geodesics at the initial point and the final position $(u_\infty, \zeta_\infty,
\bar\zeta_\infty)$ at $\cal{I}^+$.  One way to find the points of $\cal{I}^+$ which are conjugate to
the initial point $x^a_\circ$ is to find the points at $\cal{I}^+$ for which the Jacobian matrix
expressing the mapping,

\begin{equation} J = \pmatrix{\frac{\partial u_\infty}{\partial b}&\frac{\partial u_\infty}{\partial
\gamma} \cr \frac{\partial \zeta_\infty}{\partial b}&\frac{\partial \zeta_\infty}{\partial \gamma}\cr
\frac{\partial\bar\zeta_\infty}{\partial b}&\frac{\partial \bar\zeta_\infty}{\partial \gamma}\cr},
\label{J} \end{equation}

\noindent drops in rank~\cite{SNG}.  For this to occur, all three $2\times2$ determinants must be zero.

With no loss in generality, we consider the cut function of an initial point on the $-\hat z$ axis.  In
this case, the cut function is obtained by setting the final value of $l$ in Eq.~(\ref{lcplane}) to
zero, and restoring the rotational degree of freedom by setting $\gamma = \phi_\infty$.  Thus, the cut
function for an initial point on the $-\hat z$ axis is written in terms of the coordinates
$(u_\infty,~\theta_\infty,~\phi_\infty)$ as

\[ u_\infty(u_\circ, l_\circ, b) = u_\circ + \int_{l_\circ}^0~dl'~\left(\pm {{1 \pm
\sqrt{k(l')~A(l')}}\over{\sqrt{A(l')}(2l'^2~f(l'))}}\right), \]

\[ \theta_\infty (l_\circ, b) = \pi - \int_{l_\circ}^0~dl'~\left(\pm \frac {b}{\sqrt{A(l')}}\right),\]

\begin{equation} \phi_\infty = \gamma.  \end{equation}

\noindent In this case, the Jacobian matrix corresponding to Eq.~(\ref{J}) is given by

\begin{equation} J = \pmatrix{\frac{\partial u_\infty}{\partial b}& 0 \cr \frac{\partial
\theta_\infty}{\partial b}& 0\cr 0 & 1\cr}.  \label{Jbg}\end{equation}

\noindent It is clear that for the Jacobian to drop rank we must have

\begin{equation} \frac{\partial u_\infty}{\partial b} = 0,\label{dudb} \end{equation}

\noindent and

\begin{equation} q \equiv \frac{d \theta_\infty}{db} = 0, \label{dtheta} \end{equation}

\noindent A combination of numerical and analytic calculations shows that these two conditions are
satisfied simultaneously only at the cusp ridge shown in the figures, and hence the cusps on the cut
function are conjugate points to the initial point.

An alternative way of deducing the singular points is to consider the first and second derivatives of
$u_\infty$ with respect to $\theta_\infty$.  For a cusp singularity, the first derivatives are always
finite, while the second derivatives diverge.  Working parametrically in $b$, these derivatives are

\[ \frac{\partial u_\infty}{\partial \theta_\infty} = \frac{\frac{\partial u_\infty}{\partial b}}
{\frac{\partial \theta_\infty}{\partial b}}, \] \noindent and,

\begin{equation} \frac{\partial^2 u_\infty}{\partial \theta_\infty^2} = \left( \frac{\partial
\theta_\infty}{\partial b} \right)^{-2}\left[\frac{\partial^2 u_\infty}{\partial b^2} - \frac{\partial
u_\infty}{\partial \theta_\infty} \frac{\partial^2 \theta_\infty}{\partial b^2} \right].
\end{equation}

\noindent Numerical computation shows that the first derivative is finite and non-zero along the cusps
in the light cone cut.  As written, the second derivative is indeterminant along the cusps since the
term in square brackets is zero there.  Applying L'H\^ospital's rule, one sees that the second
derivative actually diverges as $q^{-1}$.

As a final note on the cusp singularities, we list the behavior of several important quantities in the
Null Surface Formulation as one approaches the cusps along the cut function.  All of these quantities
are computed as derivatives of the cut function with respect to the complex stereographic coordinates
$\zeta$ and $\bar\zeta$, and have been computed in this model parametrically for initial points along
the $-\hat z$ axis.  These derivatives are denoted by $\mbox \dh$ and $\bar{\mbox \dh}$~\cite{eth}.  In
terms of $q$, which approaches zero as one approaches the cusp, the behavior of some important
quantities of interest are listed below for $u_\infty~=~Z(x_\circ^a, \zeta, \bar\zeta)$.

\begin{center} \begin{tabular}{ccc} Quantity & Behavior \\ \hline $\omega = \mbox \dh Z$ & regular \\
$\Lambda = {\mbox \dh}^2 Z$ & $q^{-1}$ \\ $R = \bar{\mbox \dh}\mbox \dh Z$ & $q^{-1}$ \\
$|\Lambda,_{1}| = |\frac{d \Lambda}{dR}|$ & $1$ \\ $\Omega^2 = g^{ab}Z,_a\bar{\mbox \dh}\mbox \dh Z,_b$
& $q^{-2}$ \end{tabular} \end{center}

\section{Gravitational Lensing Equations}

An important goal of gravitational lensing theory is to construct lens equations which give the
position of sources in terms of directions seen by an observer and distances to the source.  Typically,
lens equations are obtained via approximations on the kinematics of the null geodesics of the
source~\cite{Ehlers}.

Recently, a way to produce completely general, exact lensing equations has been found, and a paper is
being prepared which develops gravitational lensing theory from this perspective~\cite{SNG}.  Our model
provides an explicit example of such a formulation.  In this section we give exact lensing equations
for the Schwarzschild space-time with a constant density dust interior region, and show that our exact
equations reduce to standard approximate lens equations.  In the special case that the source, lens,
and observer are spatially colinear, we give an exact expression for the observation angle in an
Einstein Ring.

In Section 2, we derived the future and past light cones of an arbitrary point in the space-time.
Recall that a lens equation should express the location of the source in terms of some ``distance''
from the observer, and the directions which the observer views the geodesics on the past light cone.
The equations of the past light cone, Eqs.~(\ref{lightcone}), are such a set of equations.  In these
equations, the observed directions are given by the parameters $(b, \gamma)$, and $l$ gives the
``distance.'' Hence, exact lens equations for our model are

\[ \zeta(l_\circ, \zeta_\circ, \bar\zeta_\circ, l, b^\ast, \gamma^\ast) = \left(\frac{\zeta_\circ}{\bar
\zeta_\circ}\right)^{\frac{1}{2}}\left(\frac{e^{\frac{i}{2}\gamma^\ast} \cot{\frac{\Theta(l, l_\circ,
b^\ast)}{2}} + \sqrt{\zeta_\circ\bar\zeta_\circ}
e^{-\frac{i}{2}\gamma^\ast}}{-\sqrt{\zeta_\circ\bar\zeta_\circ} e^{\frac{i}{2}\gamma^\ast}
\cot{\frac{\Theta(l, l_\circ, b^\ast)}{2}} + e^{-\frac{i}{2}\gamma^\ast}}\right), \]

\begin{equation} \bar\zeta(l_\circ, \zeta_\circ, \bar\zeta_\circ, l, b^\ast, \gamma^\ast) =
\left(\frac{\bar\zeta_\circ}{\zeta_\circ}\right)^{\frac{1}{2}}\left(\frac{e^{-\frac{i}{2}\gamma ^ \ast}
\cot{\frac{\Theta(l, l_\circ, b^\ast)}{2}} + \sqrt{\zeta_\circ\bar\zeta_\circ}
e^{\frac{i}{2}\gamma^\ast}}{-\sqrt{\zeta_\circ\bar\zeta_\circ} e^{\frac{-i}{2}\gamma^\ast}
\cot{\frac{\Theta(l, l_\circ, b^\ast)}{2}} + e^{\frac{i}{2}\gamma^\ast}}\right),
\label{lens1}\end{equation}

\noindent with

\[ \Theta(l, l_\circ, b^\ast) = \pi - \int_{l_\circ}^l~dl'~\left(\pm \frac
{b^\ast~\sqrt{k(l')}}{\sqrt{1 - {b^\ast}^2~l'^2~f(l')}}\right).  \]

In these lens equations, the spatial location of the source is the point $(l, \zeta, \bar\zeta)$.  For
an observer at the point $(l_\circ, \zeta_\circ, \bar\zeta_\circ)$, the observed directions of the
geodesic on the past null cone are given by the particular values of $(b, \gamma)$ which connect the
source and the observer, denoted as $(b^\ast, \gamma^\ast)$.  There may, in fact, be more than one set
of values for $(b^\ast, \gamma^\ast)$, as the process of focusing may produce more than one ``image.''

Due to spherical symmetry, any observer may be considered as lying on the $-\hat z$ axis, and the
source may be taken as lying in the $\hat x$-$\hat z$ plane.  We are interested in the case where the
lens is situated between the observer and the source, and when the rays do not pass through the
interior region of the star.  In this case, the lens equations, Eqs.~(\ref{lens1}), reduce to the
single equation for $\Theta$ which was found in Eq.~(\ref{lcplane}):

\begin{equation} \Theta (l_\circ, l, b^\ast) =\pi - \int_{l_\circ}^l~dl'~\left(\pm \frac
{b^\ast}{\sqrt{1 - {b^\ast}^2~l'^2~f(l')}}\right).  \label{lenstheta} \end{equation}

\noindent This lens equation specifies the location of the source in terms of the observed direction of
the geodesic, $b^\ast$, and a ``distance,'' $l$, to the source.

In Appendix A, we find the relationship between the angle at which a null geodesic crosses the $\hat z$
axis, the parameter $b$, and a position $l$.  In the lensing case, this ``observation angle,'' denoted
by $\psi_{obs}$, is related to the observer position, $l_\circ$, and the observed direction, $b^\ast$,
by

\begin{equation} b^\ast = \frac{\sin{\psi_{obs}}}{l_\circ~\sqrt{f(l_\circ)}}.  \label{bast}
\end{equation}

\noindent By replacing $b^\ast$ by $\psi_{obs}$ in the lens equation, Eq.~(\ref{lenstheta}), the lens
equation takes a more conventional form, where the direction parameter is the actual observation angle:

\begin{equation} \Theta (l_\circ, l, b^\ast) =\pi - \frac{\sin{\psi_{obs}}}{l_\circ~\sqrt{f(l_\circ)}}
~\int_{l_\circ}^l~dl'~\left(\pm \frac {1}{\sqrt{1 -
\frac{\sin^2{\psi_{obs}}~l'^2~f(l')}{l_\circ^2~f(l_\circ)}}}\right).  \label{lens} \end{equation}

A typical approximate lens equation for the Schwarzschild model~\cite{Ehlers} is

\begin{equation} \beta = \psi - \frac{2R_S~D_{LS}}{D_L~D_S~\psi}.\label{approxlens} \end{equation}

\begin{figure} \begin{center} \scalebox{1.0}{\includegraphics{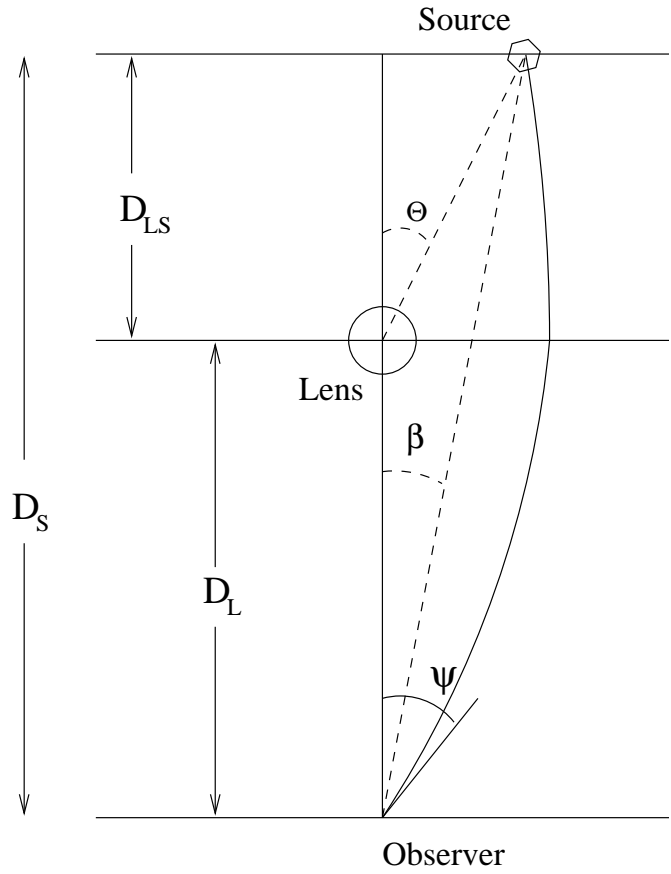}} \caption{The schematic
representation of the path of a geodesic observed in gravitational lensing.  Distances between the lens
and observer, lens and source, and observer and source are shown along with the observation angle,
$\psi$.}  \label{truelens} \end{center} \end{figure}

\noindent In this approximation, $\beta$ is the Euclidean angle between the source and the center of
the space-time, and $R_S = 2M$ is the Schwarzschild radius.  The Euclidean distances between the source
and lens, source and observer, and lens and observer, are given by $D_{LS}$, $D_S$, and $D_L$
respectively.  Figure~\ref{truelens} shows the case under consideration.  We now show that our lens
equation, Eq.~(\ref{lens}) or Eq.~(\ref{lenstheta}), reduces to the approximate formula,
Eq.~(\ref{approxlens}), under appropriate approximations.

Taking into account the correct signs for incoming and outgoing rays, the right hand side of
Eq.~(\ref{lenstheta}) can be written as

\begin{equation} \Theta = \pi - \Delta(M, b^\ast, l_\circ, l), \label{working} \end{equation}

\noindent where

\begin{eqnarray} \Delta &= &2~\int_{l_\circ}^{l_b}~\frac{ b^\ast~dl'}{\sqrt{2\sqrt{2}M~{b^\ast}^2~l'^3
- {b^\ast}^2~l'^2 +1}} \nonumber \\ & & {} + \int_{l}^{l_\circ}~\frac{
b^\ast~dl'}{\sqrt{2\sqrt{2}M~{b^\ast}^2~l'^3 - {b^\ast}^2~l'^2 +1}}.\label{fullpsi} \end{eqnarray}

\noindent Here, $l_\circ$ is the position of the observer, $l$ is the position of the source, and $l_b$
is the value of $l$ for which the geodesic comes closest to the lens, attained when $\dot l = 0$.  The
maximum $l$ value, $l_b$, is, from Eq.~(\ref{lmax}), the solution of the equation

\begin{equation} 1 - {b^\ast}^2~l_b^2 + 2\sqrt{2}M~{b^\ast}^2~l_b^3 = 0.  \label{lbeqn} \end{equation}

For convenience, we assume that the source is closer to the lens than the observer, so that $l >
l_\circ$.  To proceed, we assume that the dimensionless quantities $M~l \equiv \epsilon$ and $M~l_\circ
< \epsilon$ are small and make a Taylor series expansion of $\Delta$ in terms of $\epsilon$:

\[ \Delta(\epsilon, b^\ast, l_\circ, l) = \Delta(\epsilon = 0, b^\ast, l_\circ, l) + \epsilon
\left[\frac{\partial \Delta}{\partial \epsilon}\right]_{\epsilon = 0} \equiv \Delta_\circ + \Delta_1.
\]

To compute $\Delta_\circ$, we evaluate Eq.~(\ref{fullpsi}) at $\epsilon=M=0$.  This implies that $l_b =
1/b^\ast$ from Eq.~(\ref{lbeqn}).  In this case the integrals are all trigonometric integrals, and we
have

\[ \Delta_\circ = \pi - \arcsin{(b^\ast~l_\circ)} - \arcsin{(b^\ast~l)}.  \]

\noindent Using Eq.~(\ref{bast}) to express $\Delta_\circ$ in terms of $\psi_{obs}$, and making a small
angle approximation, $\Delta_\circ$ is given by

\begin{equation} \Delta_\circ = \pi - \psi_{obs} - \frac{\psi_{obs}~l}{l_\circ}.  \end{equation}

The first order term is given by

\begin{equation} \Delta_1 = \epsilon~\left[\frac{d}{d\epsilon}\right]_{\epsilon =0}
\left(2~\int_{l_\circ}^{l_b}~\frac{ b^\ast\sqrt{l}~dl'}{A} + \int_{l}^{l_\circ}~\frac{
b^\ast\sqrt{l}~dl'}{A} \right), \end{equation}

\noindent where

\[ A = \sqrt{2\sqrt{2}\epsilon~{b^\ast}^2~l'^3 - {b^\ast}^2~l~l'^2 + l}.  \]

\noindent The derivative acts on both the $\epsilon$ dependence in the integrals and in the upper limit
$l_b$, and care must be taken so that there is a cancellation of two divergent pieces which appear.
Using a small angle expansion in $\psi_{obs}$, the first order correction to $\Theta$ is

\begin{equation} \Delta_1 = \frac{4 \sqrt{2}M~l_\circ}{\psi_{obs}}.  \end{equation}

Inserting the forms of $\Delta_\circ$ and $\Delta_1$ into Eq.~(\ref{working}) gives

\begin{equation} \Theta = \psi_{obs}(1 + \frac{l}{l_\circ}) - \frac{2\sqrt{2}R_S~l_\circ}{\psi_{obs}},
\label{almost} \end{equation}

\noindent where $R_S = 2M$ is the Schwarzschild radius.

To lowest order, the physical distances in Figure \ref{truelens} are the inverse coordinate distances,

\[l \approx \frac{1}{\sqrt{2}D_{LS}} \quad\quad\quad l_\circ \approx \frac{1}{\sqrt{2}D_L}, \]

\noindent and from Euclidean geometry, $\Theta$ is related to $\beta$ by

\[ \Theta = \frac{\beta~D_S}{D_{LS}}.  \]

\noindent Using these relationships in Eq.~(\ref{almost}) and rearranging gives an approximate lens
equation

\[ \beta = \frac{D_{LS} + D_L}{D_S}\psi_{obs} - \frac{2~R_S~D_{LS}}{D_S~D_L~\psi_{obs}} \label{there}
\]

\noindent which is the standard result when $D_L + D_{LS} = D_S$:

\begin{equation} \beta = \psi_{obs} - \frac{2~R_S~D_{LS}}{D_S~D_L}\frac{1}{\psi_{obs}} .
\end{equation}

As a special case, we consider the Einstein Rings, an early prediction of ``pre''-General Relativity
only recently observed.  If the source lies along the $+\hat z$ axis, directly opposite the lens from
the observer at $\beta = \Theta = 0$, the observer sees the image as a circular ring surrounding the
lens, or an Einstein Ring.  In this special case, the lens equation, Eq.~(\ref{lens}), is an implicit
equation for the exact observation angle for the ring:

\begin{eqnarray} \pi &=& \frac{\sin{\psi_{obs}}}{l_\circ~\sqrt{f(l_\circ)}}
~\int_{l_\circ}^{l_b}~dl'~\left( \frac {1}{\sqrt{1 -
\frac{\sin^2{\psi_{obs}}~l'^2~f(l')}{l_\circ^2~f(l_\circ)}}}\right)\nonumber \\ & & {} -
\frac{\sin{\psi_{obs}}}{l_\circ~\sqrt{f(l_\circ)}} ~\int_{l_b}^{l}~dl'~\left( \frac {1}{\sqrt{1 -
\frac{\sin^2{\psi_{obs}}~l'^2~f(l')}{l_\circ^2~f(l_\circ)}}}\right), \end{eqnarray}

\noindent where $l_b$ is the postive root of Eq.~(\ref{lbeqn}) or the positive root of

\begin{equation} {(\sin \psi_{obs} )^{2}}~l_b^{2}(1-2\sqrt{2}Ml_b)-l_{\circ}^{2}~(1-2\sqrt{2}
Ml_{\circ})=0.\end{equation}

In terms of the future light cone of the source, an observer who sees the Einstein Ring is situated
along the crossover line in Figure~\ref{lightconepic}.  Points on this line are conjugate to the
initial point, and the light cone has unstable singularities there.  The crossover point in the cut
function represents a limiting ``Einstein Ring'' at infinity, but the actual observation angle for this
ring is zero, so that the ring is not observable from infinity.

\section{Pseudo-Minkowski Coordinates}

As a final application of the cut function, we show that the so called pseudo-Minkowski
coordinates~\cite{simo3} form a well defined, global coordinate system for the Schwarzschild space-time
with a constant density dust interior.  In this section, we use the complex stereographic angles
$(\zeta, \bar\zeta)$ as coordinates on the sphere.

The pseudo-Minkowski coordinates are defined by integrals over the sphere at infinity of the cut
function weighted against the first four $Y_{lm}$,

\begin{equation} x_{l,m} = \int_{S^2}~Z(x^a_\circ, \zeta)~\bar{Y}_{l,m}(\zeta)~dS^2 \equiv
f_{l,m}(x^a_\circ), \quad\quad (l = 0, 1) \label{def1} \end{equation}

\noindent where

\[ dS^2 = \frac{2}{i}~\frac{d\zeta~\wedge~d\bar\zeta}{(1 + \zeta \bar\zeta)^2} \]

\noindent is the volume element on the sphere of the null generators of $\cal{I}^+$.  There is a
conceptual problem with the definition of the pseudo-Minkowski coordinates as stated in
Eq.~(\ref{def1}).  Namely, the definition is ambiguous because the cut function $u =
Z(x^a_\circ,\zeta)$, is, in general, not single valued at $\cal{I}^+$, and so one does not know which
portion of the cut to integrate over.

The ambiguity is resolved by using the light cone structure to pull the integral back to the sphere of
initial null directions at the initial point $x^a_\circ$.  To pull the integral back, we must have a
function,

\begin{equation} \zeta = \zeta(x^a_\circ, \eta, \bar \eta), \label{zetaeta} \end{equation}

\noindent which relates the final angular positions at $\cal{I}^+$, the $(\zeta, \bar\zeta)$ to the
initial direction of the geodesic, $(\eta, \bar\eta)$ at the initial point.  Given a function of the
form of Eq.~(\ref{zetaeta}), we can form the determinant of the Jacobian matrix,

\begin{equation} |J| = \frac{\partial \zeta}{\partial \eta}\frac{\partial \bar\zeta}{\partial \bar\eta}
- \frac{\partial \zeta}{\partial \bar\eta}\frac{\partial \bar\zeta}{\partial \eta}, \end{equation}

\noindent and transform the integral from an integral over the sphere at null infinity into an integral
over the sphere of initial directions:

\begin{equation} x_{l,m} = \int_{S^2_\circ}~Z(x^a_\circ, \zeta(x^a_\circ, \eta))
~\bar{Y}_{l,m}(\zeta(x^a_\circ, \eta))~|J|~dS_\circ^2 = f_{lm}(x^a_\circ)\quad\quad (l = 0, 1).
\label{defxlm} \end{equation}

\noindent This integral defines the pseudo-Minkowski coordinates.

We would like to show that the pseudo-Minkowski coordinates form a good coordinate system by showing
that the Jacobian of the coordinate transformation defined by Eq.~(\ref{defxlm}),

\begin{equation} x_{lm} =f_{lm}(x^a_\circ), \end{equation}

\noindent is non-zero.

Due to the spherical symmetry of the space-time and the fact that the $x_{1,m}~=~(x_{1,1}, x_{1,0},
x_{1,-1})$ transforms as an $O(3)$ vector under space-time rotations, we can conclude that the
functional form of the pseudo-Minkowski coordinates must be

\[ x_{1, -1} = X - iY = f(u_\circ, l_\circ, b, \gamma)\sin{\theta}~e^{-i\phi},\]

\[x_{1, 0} = Z = f(u_\circ, l_\circ, b, \gamma)\cos{\theta}, \]

\[ x_{1, 1} = X + iY = f(u_\circ, l_\circ, b, \gamma)\sin{\theta}~e^{i\phi},\]

\begin{equation} x_{0, 0} = T = g(u_\circ, l_\circ, b, \gamma).  \label{form} \end{equation}

\noindent To test the non-vanishing of the Jacobian, all we need to do is to take a point of the
$(u_\circ, l_\circ)$ plane, for example $\phi_\circ = 0$, and $\theta_\circ = \pi$, and check the
transformation

\begin{equation} x_{0,0} = g(u_\circ, l_\circ, b, \gamma), \quad\quad x_{1,0} = f(u_\circ, l_\circ, b,
\gamma), \end{equation}

\noindent since this part of the coordinate transformation represents the ``non-rotational'' part.  The
determinant, $D$, of interest is given by

\begin{equation} D = \frac{\partial x_{00}}{\partial u_\circ}\frac{\partial x_{10}}{\partial l_\circ} -
\frac{\partial x_{00}}{\partial l_\circ}\frac{\partial x_{10}}{\partial u_\circ}.  \end{equation}

For points along the $-\hat z$ axis, using $b$ and $\gamma$ as the initial parameters and the Jacobian
expressing their relationship to the angles $(\zeta, \bar\zeta)$ found in Appendix B, the
pseudo-Minkowski coordinates are

\begin{equation} x_{l,m} = \int~db~\wedge~d\gamma~\sin{\theta_\infty(l_\circ, b)}\frac{\partial
\theta_\infty}{\partial b}~u(u_\circ, l_\circ, b)~\bar{Y}_{l,m}(\zeta(b, \gamma)),\quad\quad (l = 0,1)
\label{xlm} \end{equation}

\noindent where the range in $\gamma$ is zero to $2\pi$ and the range in $b$ runs fully over both
sheets of solutions.  The integration over $\gamma$ does not cause any trouble for any of the
integrals.  At first glance, the convergence of the integration over $b$ is not clear, due to
divergences in term $\partial \theta_\infty / \partial b$ as $b$ approaches its maximum value,

\[ b_m = \frac{1}{l_\circ~\sqrt{f(l_\circ)}}.  \]

\noindent These divergences are all of order $(b_m - b)^{-\beta}$ with $\beta < 1$, which ensures that
the $b$ integral also converges.  The derivatives in question can be written as:

\[ \frac{\partial x_{00}}{\partial u_\circ} = \sqrt{\pi}~\int~db~ \sin{\theta_\infty(l_\circ, b)}
~\frac{\partial \theta_\infty}{\partial b}, \]

\[ \frac{\partial x_{00}}{\partial l_\circ} = \sqrt{\pi}~\frac{\partial}{\partial l_\circ}~\int~db
~u(l_\circ, u_\circ, b)~ \sin{\theta_\infty(l_\circ, b)} ~\frac{\partial \theta_\infty}{\partial b}, \]

\[ \frac{\partial x_{10}}{\partial u_\circ} = \sqrt{3\pi}~\int~db ~\sin{\theta_\infty(l_\circ, b)}~
\cos{\theta_\infty(l_\circ, b)}~\frac{\partial \theta_\infty}{\partial b}, \]

\begin{eqnarray} \frac{\partial x_{10}}{\partial l_\circ} &= &\sqrt{3\pi}~\frac{\partial}{\partial
l_\circ}~\int~db~ u(l_\circ, u_\circ, b)~\sin{\theta_\infty(l_\circ, b)} ~\cos{\theta_\infty(l_\circ,
b)}~\frac{\partial \theta_\infty}{\partial b} \nonumber\\ & =&
\frac{\sqrt{3\pi}}{2}~\frac{\partial}{\partial l_\circ}~\int~db~ u(l_\circ, u_\circ,
b)~\frac{d}{db}\left( (\sin{\theta_\infty(l_\circ, b)})^2\right).  \end{eqnarray}

The integrals are defined piecewise along the various segments of $u(l_\circ, u_\circ, b)$ and
$\theta_\infty(l_\circ, b)$.  The range in $b$ runs from $b =0$, when the null geodesic is radially
outgoing, and hence $\theta_\infty(l_\circ, b=0) = \pi$, to a maximum value to $b = b_m$, and, on the
second sheet, back down to $b = 0$ for radially ingoing rays, where $\theta_\infty(l_\circ, b=0) = 0$.
The first integral is easily performed:

\begin{eqnarray} \frac{\partial x_{00}}{\partial u_\circ} &=& \sqrt{\pi}~\int~db
~\sin{\theta_\infty(l_\circ, b)} ~\frac{\partial \theta_\infty}{\partial b} = \sqrt{\pi}~\int~db
~\frac{d}{db}~(-\cos{\theta_\infty}) \nonumber \\ &=& -\sqrt{\pi}(\cos{0} - \cos{\pi}) = -2\sqrt{\pi}.
\end{eqnarray}

\noindent Likewise,

\begin{eqnarray} \frac{\partial x_{10}}{\partial u_\circ} &=& \sqrt{3\pi}~\int~db~
\sin{\theta_\infty(l_\circ, b)} ~\cos{\theta_\infty(l_\circ, b)}~\frac{\partial \theta_\infty}{\partial
b} \nonumber\\ &=& \sqrt{3\pi}~\int~db ~\frac{d}{db}\left(\frac{1}{2}(\sin{\theta_\infty})^2\right)
\nonumber \\ &=& \frac{\sqrt{3\pi}}{2}\left((\sin{0})^2 - (\sin{\pi})^2\right) = 0.  \end{eqnarray}

\noindent Therefore, when the initial point lies along the $-\hat z$ axis, the Jacobian of the
transformation simplifies to

\begin{eqnarray} D &=& \frac{\partial x_{00}}{\partial u_\circ}\frac{\partial x_{10}}{\partial l_\circ}
- \frac{\partial x_{00}}{\partial l_\circ}\frac{\partial x_{10}}{\partial u_\circ} =
-2\sqrt{\pi}\frac{\partial x_{10}}{\partial l_\circ} \nonumber \\ &=&
-\sqrt{3}\pi~\frac{\partial}{\partial l_\circ}~\int~db~ u(l_\circ, u_\circ,
b)\frac{d}{db}\left((\sin{\theta_\infty(l_\circ, b)})^2\right), \nonumber\end{eqnarray}

\noindent or finally

\begin{equation} D = -\sqrt{3}\pi \frac{\partial}{\partial l_\circ}~{\cal{I}}(l_\circ), \end{equation}

\noindent with

\begin{equation} {\cal{I}}(l_\circ) = \int~db~ u(l_\circ, u_\circ,
b)\frac{d}{db}\left((\sin{\theta_\infty(l_\circ, b)})^2\right).  \end{equation}

\begin{figure} \begin{center} \scalebox{1.0}{\includegraphics{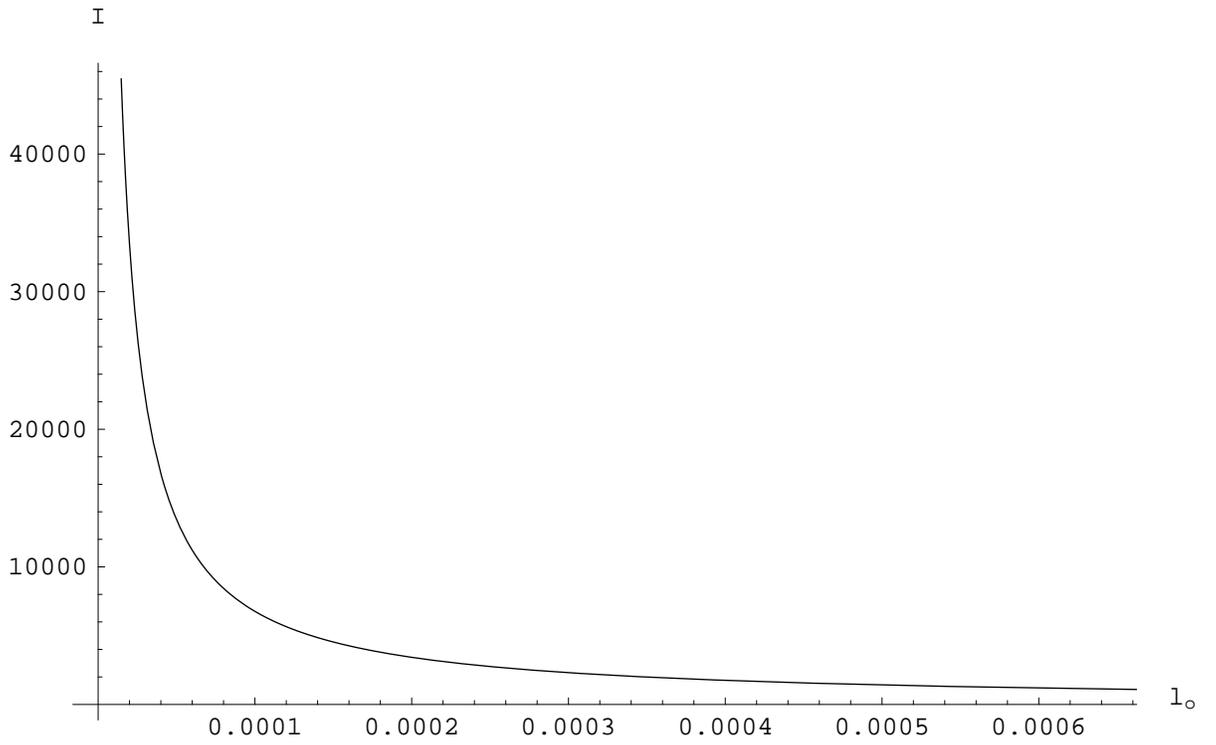}}\caption{A plot of the
integral, ${\cal{I}}(l_\circ)$, as a function of the initial radial position shows that there are no extrema. The integral has a finite derivative for all points except $l_\circ = 0$, which is not in the physical space-time.} \label{jacadata} \end{center} \end{figure}

Thus, to determine if the pseudo-Minkowski coordinates are a good coordinate system, we only have to
show that the integral ${\cal{I}}(l_\circ)$ has no extremum as the initial radial coordinate parameter,
$l_\circ$, is varied.  An extensive numerical calculation shows that there are no extremum to this
integral, whose values are plotted for many initial positions in Figure~\ref{jacadata}.  In fact, the
integral is a constantly decreasing function, whose derivative is finite at all points $l_\circ$ except
$l_\circ = 0$, which corresponds to spatial infinity.  Since spatial infinity is not a point in the
space-time, we claim that the determinant, $D$, has a finite positive value for all initial positions.

We have shown that the Jacobian of the transformation between the $(u_\circ, l_\circ)$ and $(x_{0,0},
x_{1,0})$ portions of the total coordinate transformation, \[ x_{l,m}~=~f_{l,m}(x^a_\circ), \]

\noindent is non-zero.  Using the spherical symmetry of the space-time and the inherent transformation
properties of the $x_{l,m}$, we can claim that the entire transformation is non-singular, or that that
the pseudo-Minkowski coordinates are a good coordinate system of Schwarzschild space-time with a
constant density dust interior.

\appendix

\section{The Initial Direction and the Parameter b}

The parameter $b$, which arose as a constant of integration when integrating the null geodesic
equations, parametrized the initial direction of the geodesic.  In this appendix, we choose the motion
of the geodesic to remain in the $\hat x$-$\hat z$ plane and the initial point to lie on the $-\hat z$
axis.  The initial direction of a geodesic is captured by giving an angle, $\psi$, between the spatial
part of the directed tangent vector to the geodesic and the $\hat z$ axis.  We are interested in
determining the relationship between the angle $\psi$ and the parameter $b$.

From Eq.~(\ref{lcplane}), the coordinates of a null geodesic restricted to the $\hat x$-$\hat z$ plane
were given in terms of $l$ by

 \[ u = u_\circ + (-1)^\epsilon \int_{l_\circ}^l~dl' \left( \pm {{1 \pm
\sqrt{k(l')~A(l')}}\over{\sqrt{A(l')}(2l'^2~f(l'))}}\right), \] \[ l = l,\] \[ \theta =\pi -
\int_{l_\circ}^l~dl'~\left(\pm \frac {b}{\sqrt{A(l')}}\right), \] \begin{equation} \phi = 0,
\label{eq2} \end{equation}

\noindent with \[ A(l') = \frac{1 - b^2~{l'}^2~f(l')}{f(l')~h(l')}.  \]

\noindent Up to rescaling, the (null) tangent vector to this geodesic is

\begin{equation} L^a = \left(\frac{du}{dl}, \frac{dl}{dl}, \frac{d\theta}{dl}, \frac{d\phi}{dl}\right)
= \left(\frac{du}{dl}, 1, \frac{d\theta}{dl}, 0\right).  \label{L} \end{equation}

Since both the measure of angles and the length of a null vector are independent of conformal factors,
any conformally related metric may be used to compute them.  In what follows we use the physical metric
of our model, which in the coordinates $(t, l= 1/(\sqrt{2}r), \theta, \phi)$ is

\begin{equation} ds^2 = f(l) dt^2 - \frac{h(l)}{2l^4}dl^2 - \frac{1}{2l^2}d\Omega^2 = f(l) dt^2 -
g_{ij}dx^i dx^j, \end{equation}

\noindent where $f(l)$ and $h(l)$ are the coefficients of the metric given in Eq.~(\ref{fandh}) and
$g_{ij}$ is a spatial metric.  The spatial part of null vector, Eq.~(\ref{L}), normalized in the
physical metric, is the three vector

\begin{equation} \hat{L}^i = \frac{1}{| L |}\left( 1, \frac{d\theta}{dl}, 0 \right), \end{equation}

\noindent with

\[ | L | = \sqrt{\frac{h(l)}{2l^4} + \frac{1}{2l^2}\left(\frac{d\theta}{dl}\right)^2}.  \]

\noindent The value of the derivative $d\theta / dl$ is determined using Eq.~(\ref{eq2}):

\begin{equation} \left(\frac{d\theta}{dl}\right)^2 = \frac{ f(l)~h(l)~ b^2 }{1 - b^2~l^2~f(l)}.
\end{equation}

\noindent A unit spatial vector pointing in the radial direction is given by

\[ \hat{r}^i = \left( \frac{\sqrt{2}l^2}{\sqrt{h(l)}}, 0, 0 \right).  \]

The inner product between $L^i$ and $r^i$ gives the angular direction of the geodesic, namely,

\begin{equation} g_{ij}\hat{r}^i\hat{L}^j = \cos{\psi}.  \label{tree}\end{equation}

\noindent After some algebra, Eq.~(\ref{tree}) can be solved for $b$, giving our desired result

\begin{equation} b = \frac{\sin{\psi}}{l~\sqrt{f(l)}}.  \label{delta} \end{equation}

The range in $b$ at the initial point is determined by Eq.~(\ref{delta}).  The parameter ranges from $b
= 0$, corresponding to radially outgoing rays when $\psi = 0$, to a maximum value, $b~=~b_m$, where
$\psi = \pi/2$, back down to $b~=~0$ where $\psi = \pi$, and again the geodesic travels radially.
Either $b$ or $\psi$ may be used to parametrize the initial direction of the geodesic.

\section{Full Angular Dependence of the Light Cone}

In Section 2, we integrated the null geodesics emanating from a point on the $-\hat z$ axis, restricted
to the $\hat x$-$\hat z$ plane, in terms of a parameter $l$.  The angular integrals were

\[ \phi = 0, \] \begin{equation} \Theta = \Theta (l, l_\circ, b) = \pi - \int_{l_\circ}^l~dl'~\left(\pm
\frac {b~\sqrt{k(l')}}{\sqrt{1 - b^2~l'^2~f(l')}}\right).  \label{thetaapp} \end{equation}

\noindent We want to perform a rigid rotation of this restricted solution restoring the full angular
dependence, and allowing the initial point to be at any position.

Due to spherical symmetry, the geodesic equations separate into two time/radial equations and two
angular equations.  An arbitrary solution to the angular part of the geodesic equations can be obtained
by performing a rigid rotation of the solution given in Eq.~(\ref{thetaapp}).  For such a solution, the
motion will take place in a new plane, but the angle $\Theta(l, l_\circ, b)$ will be preserved.

To perform the rotation we use complex stereographic coordinates, $(\zeta,~\bar\zeta)$, as coordinates
on the sphere defined in Eq.~(\ref{defzeta}).  In terms of $\zeta$, the solution corresponding to
Eq.~(\ref{lcplane}) is

\begin{equation} \zeta (l,l_\circ, b) = \cot{\frac{\Theta (l, l_\circ, b)}{2}} = \bar\zeta(l, l_\circ,
b).  \end{equation}

\noindent Under an $SU(2)$ rotation, $\zeta$ transforms as

\begin{equation} \zeta' = \frac{\mathsf{a} \zeta + \mathsf{b}}{\mathsf{c} \zeta + \mathsf{d}} =
\frac{\mathsf{a} \cot{\frac{\Theta(l, l_\circ, b)}{2}} + \mathsf{b}}{\mathsf{c} \cot{\frac{\Theta(l,
l_\circ, b)}{2}} + \mathsf{d}}, \label{trans} \end{equation}

\noindent where $\mathsf {a, b, c, d}$ are the Cayley-Klein parameters~\cite{Goldstein}, which can be
expressed in terms of Euler angles, $\alpha$, $\beta$, and $\gamma$ as

\[ \mathsf{a} = \sin{\frac {\alpha}{2}} e^{\frac{i}{2}(\gamma + \beta)}, \] \[ \mathsf{b} = \cos{\frac
{\alpha}{2}} e^{\frac{i}{2}(- \gamma + \beta)}, \] \[ \mathsf{c} = - \cos{\frac {\alpha}{2}}
e^{\frac{i}{2}( \gamma - \beta)}, \] \begin{equation} \mathsf{d} = \sin{\frac {\alpha}{2}}
e^{\frac{i}{2}(- \gamma - \beta)}.  \end{equation}

\noindent To determine the values of the Euler angles, we note that when $\Theta = \pi$, the geodesic
is at the initial position, $\zeta_\circ$.  From $\zeta' = \zeta_\circ$, we have:

\begin{equation} \frac{\mathsf{b}}{\mathsf{d}} = \cot{\frac{\alpha}{2}} e^{i\beta} =
\cot{\frac{\theta_\circ}{2}}e^{i\phi_\circ}.  \end{equation}

\noindent This condition fixes two of the Euler angles, $\alpha$ and $\beta$, to be
$\alpha~=~\theta_\circ$ and $\beta~=~\phi_\circ$.  Thus, in terms of the new initial point,
$\zeta_\circ$, the Cayley-Klein parameters are

\begin{eqnarray} \mathsf{a} &=& \sqrt{\frac{1}{1 + \zeta_\circ\bar\zeta_\circ}}
\left(\frac{\zeta_\circ}{\bar \zeta_\circ}\right)^{\frac{1}{4}} e^{\frac{i}{2} \gamma}, \nonumber\cr
\mathsf{b} &=& \sqrt{\frac{\zeta_\circ \bar\zeta_\circ}{1 + \zeta_\circ\bar\zeta_\circ}}
\left(\frac{\zeta_\circ}{\bar \zeta_\circ}\right)^{\frac{1}{4}} e^{-\frac{i}{2} \gamma}, \nonumber\cr
\mathsf{c} &=& -\sqrt{\frac{\zeta_\circ \bar\zeta_\circ}{1 + \zeta_\circ\bar\zeta_\circ}}
\left(\frac{\bar \zeta_\circ}{\zeta_\circ}\right)^{\frac{1}{4}} e^{\frac{i}{2} \gamma}, \nonumber\cr
\end{eqnarray} \begin{equation} \mathsf{d} = \sqrt{\frac{1}{1 + \zeta_\circ\bar\zeta_\circ}}
\left(\frac{\bar \zeta_\circ}{\zeta_\circ}\right)^{\frac{1}{4}} e^{-\frac{i}{2} \gamma}.\label{abcd}
\end{equation}

The remaining free parameter $\gamma$ gives the orientation of the plane in which the geodesic moves.
 In the case that the initial point lies on the $-\hat z$ axis, $\gamma$ is the angle $\phi$.  When the
 initial point of the geodesic is rotated to an arbitrary location, the parameter $\gamma$ acts as an
 angle about the new axis of symmetry in the system.

Our final, full solution to the angular part of the geodesic equations is obtained using
Eqs.~(\ref{abcd}) with Eq.~(\ref{trans}):

\begin{equation} \zeta(l_\circ, \zeta_\circ, \bar\zeta_\circ, l, b, \gamma) =
\left(\frac{\zeta_\circ}{\bar \zeta_\circ}\right)^{\frac{1}{2}}\left(\frac{e^{\frac{i}{2}\gamma}
\cot{\frac{\Theta(l, l_\circ, b)}{2}} + \sqrt{\zeta_\circ\bar\zeta_\circ}
e^{-\frac{i}{2}\gamma}}{-\sqrt{\zeta_\circ\bar\zeta_\circ} e^{\frac{i}{2}\gamma} \cot{\frac{\Theta(l,
l_\circ, b)}{2}} + e^{-\frac{i}{2}\gamma}}\right), \label{zetaapp}\end{equation}

\noindent where we have dropped the prime on $\zeta$.  The angular solution, $\zeta$ is a function of
the initial point $(l_\circ, \zeta_\circ, \bar\zeta_\circ)$, a parameter along the light cone $l$, and
two free parameters, $(b, \gamma)$, which span the sphere of initial null directions at the initial
point.  The dependence on $l_\circ$, $l$, and $b$ comes through $\Theta(l, l_\circ, b)$ by integral
expression

\begin{equation} \Theta(l, l_\circ, b) = \pi - \int_{l_\circ}^l~dl'~\left(\pm \frac
{b~\sqrt{k(l')}}{\sqrt{1 - b^2~l'^2~f(l')}}\right).  \end{equation}

When the value of $l$ is taken to zero, Eq.~(\ref{zetaapp}) gives the final angular location of a point
on $\cal{I}^+$ in terms of initial directions~$(b,~\gamma)$ and the initial point $x^a_\circ =
(u_\circ, l_\circ, \zeta_\circ, \bar\zeta_\circ)$.  In this case, we denote $\zeta
(l_\circ,~\zeta_\circ,~\bar\zeta_\circ,~l = 0,~b,~\gamma)$ by $\zeta_\infty (x_\circ^a,~b,~\gamma)$,
and $\Theta (l = 0,~l_\circ,~b)$ by $\theta_\infty$.  The existence of such a function provides a
mapping from the sphere of initial null directions, $(b,~\gamma)$, to the sphere of null generators at
$\cal{I}^+$.  The Jacobian matrix of the mapping is given by

\begin{equation} J = \pmatrix{\frac{\partial \zeta_\infty}{\partial b}&\frac{\partial
\zeta_\infty}{\partial \gamma}\cr \frac{\partial \bar \zeta_\infty}{\partial b}&\frac{\partial \bar
\zeta_\infty}{\partial \gamma}\cr}, \end{equation}

\noindent In Section 5, we use the determinant of this Jacobian to transform integrals over $\cal{I}^+$
to integrals over the initial null directions.

\paragraph{Acknowledgments} We would like to thank Simonetta Frittelli for suggesting that we try to
understand the Einstein Rings in our model.  This work was supported under grants Phy 97-22049 and Phy
92-05109.

\end{document}